\documentclass[%
 reprint,
 amsmath,amssymb,
 aps,
]{revtex4-2}

\usepackage{graphicx,hyperref}
\usepackage{hyperref}
\usepackage{epstopdf}
\usepackage{dcolumn}
\usepackage{bm}
\usepackage{qcircuit}
\usepackage{subfigure}
\usepackage{MnSymbol}
\usepackage{soul}
\usepackage{comment}
\usepackage{ulem}
\usepackage{color}
\usepackage{xcolor}
\usepackage{dsfont}

\usepackage{color}
\usepackage{soul}
\usepackage{ulem}

\begin{document}

\preprint{APS/123-QED}

\title{Effect of noise on quantum circuit realization of non-Hermitian time crystals}

\author{Weihua Xie}
\email{wxx180002@utdallas.edu}
\author{Michael Kolodrubetz}%
\email{mkolodru@utdallas.edu}
\affiliation{%
 Department of Physics, The University of Texas at Dallas, Richardson, Texas 75080, USA
}%


\author{Vadim Oganesyan}
	\email{vadim.oganesyan@csi.cuny.edu}
	\affiliation{Physics program and Initiative for the Theoretical Sciences, The Graduate Center, CUNY, New York, NY 10016, USA}
	\affiliation{Center for Computational Quantum Physics, Flatiron Institute, 162 5th Avenue, New York, NY 10010, USA}    
	\affiliation{Department of Physics and Astronomy, College of Staten Island, CUNY, Staten Island, NY 10314, USA}
 %


\date{\today}

\begin{abstract}
Non-Hermitian quantum dynamics lie in an intermediate regime between unitary Hamiltonian dynamics and trace-preserving non-unitary open quantum system dynamics. Given differences in the noise tolerance of unitary and non-unitary dynamics, it is interesting to consider implementing non-Hermitian dynamics  on a noisy quantum computer. 
In this paper, we do so for a
non-Hermitian Ising Floquet model whose many-body dynamics gives rise to persistent temporal oscillations, a form of time crystallinity.  In the simplest two qubit case that we consider, there is an infinitely long-lived periodic steady state at certain fine-tuned points. These oscillations remain reasonably long-lived over a range of parameters in the ideal non-Hermitean dynamics  and for the levels of noise and imperfection expected of modern day quantum devices.  Using a generalized Floquet analysis, we
show that infinitely long-lived oscillations are generically lost for arbitrarily weak values of common types of noise and compute corresponding damping rate. We perform simulations using IBM's Qiskit platform to confirm our findings; however, experiments on a real device (ibmq-lima) do not show remnants of these oscillations.
\end{abstract}

\maketitle



\section{\label{sec:intro}Introduction}

In the canonical formulation of quantum mechanics, a system's time evolution is governed by unitary dynamics. This implies that dynamics is reversible and that phase coherence is maintained throughout the process, which is a consequence of the fact that all participating degrees of freedom are considered. However, real physical systems are never truly isolated. Interactions with the environment lead to non-unitary dynamics, which can often be approximated by the Lindblad formalism \cite{10.1093/acprof:oso/9780199213900.001.0001}. Upon post-selecting to trajectories where no quantum jumps occur, the resulting dynamics is generated by a non-Hermitian Hamiltonian. 
Non-Hermitian systems allow unique features such as unconventional band topology \cite{RevModPhys.93.015005}, skin effects \cite{PhysRevLett.121.086803}, quantum geometry \cite{PhysRevA.86.064104}\cite{Doppler2016}\cite{Xu2016}, and many others \cite{PhysRevA.97.032109,PhysRevLett.89.270401,Berry_2011,PhysRevLett.103.123601,PhysRevLett.104.054102,doi:10.1080/00018732.2021.1876991,annurev:/content/journals/10.1146/annurev-conmatphys-040521-033133}.
Furthermore, experiments have increasingly been able to reach non-Hermitian regimes and realize some of these phenomena in practice \cite{Ruter2010,Li2019,doi:10.1126/science.aar7709,Gao2015,Zhang2017,Naghiloo2019}. Given the increasing importance of non-Hermitian dynamics and their emergent phases of matter in non-equilibrium quantum physics, it is crucial to better understand how they can be realized efficiently on quantum devices and how susceptible the dynamical features are to undesired noise.

Inspired by the results of Basu et al.~\cite{PhysRevResearch.4.013018}, we will focus on a non-Hermitian realization of a time crystal. As originally proposed by Wilczek, time crystals are systems that form repeating patterns in time, spontaneously breaking time-translation symmetry \cite{PhysRevLett.109.160401}. While no-go theorems prevent their formation in equilibrium states of Hamiltonian systems \cite{PhysRevLett.114.251603}, it was soon realized that discrete time crystals may be realized by time-periodic (Floquet) drive \cite{PhysRevB.93.245146,PhysRevLett.117.090402,PhysRevLett.120.180603,PhysRevLett.120.215301,Zhang2017,Choi2017ObservationOD,Mi2022,PhysRevLett.116.250401}. In non-Hermitian Floquet systems, Ref.~\cite{PhysRevResearch.4.013018} showed that a time crystalline phase with quasi-long-range order and continuously varying period exists. In particular, the authors considered a discrete-complex-time generalization of the classical-to-quantum mapping for the Ising model \cite{beichert2013correlations}, which maps the partition function of a classical 2D Ising model to a one-dimensional quantum circuit whose non-unitary gates are generated by non-Hermitian transverse-field Ising terms. A similar model was studied in \cite{PhysRevResearch.6.013131}, revealing steady phases with robust edge modes, spatiotemporal long-range order, and transitions.

To realize this time crystal in practice, one of the most natural systems to consider is a programmable gate-based quantum computer. It is relatively straightforward to map the non-Hermitian dynamics into a quantum circuit with measurement and post-selection, but in practice the quantum computer will also have undesired noise due to interactions with its environment. Late-time unitary dynamics is in general destroyed by such noise, while non-unitary dynamics -- such as Lindblad dynamics of open quantum systems -- can be much more robust \cite{Sieberer_2016}. Non-Hermitian Floquet systems sit in between these two limiting cases and therefore it is unclear how they will respond to noise. The goal of this work will be to understand this noise tolerance for a few-qubit version of the non-Hermitian time crystal.

The remainder of this paper proceeds as follows. In section \ref{s:model}, we illustrate the quantum circuit that realizes a time crystal, along with the algorithm used to obtain two-time correlation functions. In section \ref{results}, we present data from a real device, ibmq-lima, and then model it using simulated noise. In section \ref{s:result_anlytical}, we analyze the results of the open system Floquet model. Finally, in section \ref{s:discussion}, we conclude with a discussion of generality of these results and their extension to the many-body setting.

\section{Model definition and simulation}
\label{s:model}

The system that we consider starts from the same model as in \cite{PhysRevResearch.4.013018}, namely a non-Hermitian Floquet version of the transverse-field Ising model in which each Floquet cycle consists of first applying the bond terms and then applying the field terms:
\begin{equation}
\begin{aligned}\label{VW}
    \mathcal{U} &= \mathcal{V}\mathcal{W}\\
    \mathcal{W} &= W_{1,2} W_{2,3} \cdots W_{L-1,L} = \exp{\left(J \sum_{j=1}^{L-1} \sigma_j^z \sigma_{j+1}^z\right)}\\
    \mathcal{V} &= V_1 V_2 ... V_L = \exp{\left(h \sum_{j=1}^L \sigma_j^x \right)}
\end{aligned}
\end{equation}
Importantly, both the field and interaction terms are non-Hermitian. We therefore break the coupling constants into real (non-unitary) and imaginary (unitary) parts: $h=h_R+ih_I$ and $J=J_R+iJ_I$. 

As the system time evolves from an arbitrary initial state, it quickly approaches a Floquet steady state which is similar (stroboscopically) to the ground state of a Hermitian Hamiltonian. Nevertheless, in the thermodynamic limit ($L \to \infty$), this system can order in interesting ways with uniform and non-uniform (modulated) steady state phases: paramagnet, ferromagnet,  spin-density wave, and time crystal \cite{PhysRevResearch.4.013018}. The time crystal, in particular, comes from a degeneracy in the decay rates of two single-particle Majorana modes at momenta $\pm k^\ast$. The imaginary parts of the eigenvalues are finite, resulting in persistent oscillations that display quasi-long-range order in both time and space. Linearizing around these points leads to a Fermi-surface-like dispersion for the imaginary part of the energy, enabling power-law decay of 2-point correlations \footnote{Unlike the most general non-Hermitian problem, where long-time behavior, e.g. presence of non-exponential decays, need not follow the eigenvalues but may be induced by singular eigenvector overlaps \cite{PhysRevResearch.4.013018}, the problems considered here can always be represented with \emph{symmetric} evolution operators, which guarantees that left and right eigenvectors are the same (as in unitary problems), thus making analysis of spectra sufficient.}.

\begin{figure}
  \includegraphics[width=\columnwidth]{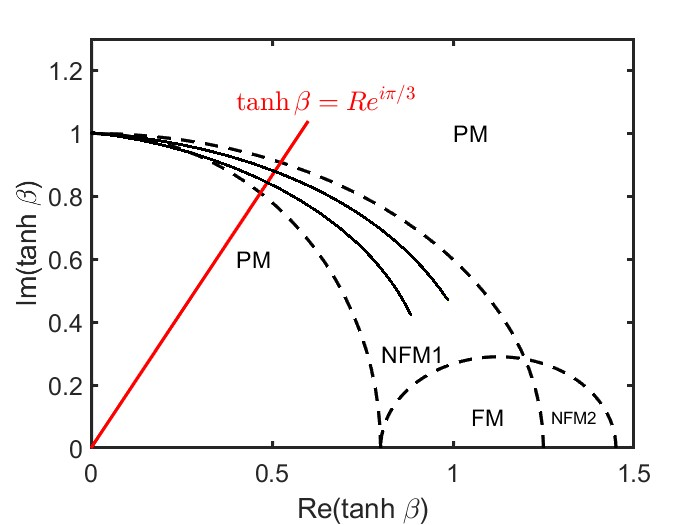}
  \caption{Sketch of the $L\to \infty$ phase diagram for our model in the $\tanh \beta$ plane, with phase transitions shown as dashed black lines (cf.~\cite{PhysRevResearch.4.013018}). The solid red line indicates the radial cut we consider throughout this text. Black solid lines show  approximate locations of steady state degeneracies for finite $L$ (here $L=2$).}
  \label{fig:phase_diagram}
\end{figure}

For finite systems, discretization of the momentum leads instead to degeneracies at fine-tuned parameter values which have an imaginary gap to the next excited state, leading to infinitely long-lived oscillations at these fine-tuned points. For this paper, we study the specific case of $L=2$ (two system qubits) \footnote{While first considered in Ref. \cite{beichert2013correlations}, that work did not appreciate the significance of space time anisotropy} with other parameters defined through analogy with the Ising partition function at complex temperature as 
\begin{equation*}
J = \beta J_y,~\tanh{h} = \exp{(-2\beta J_x)}\mbox{ with }J_x = 1,~J_y = 0.1.    
\end{equation*}
We consider a particular slice through the complex $\tanh \beta$ plane defined by 
\begin{equation}
    \tanh \beta = R \exp \left( i \pi / 3 \right).
    \label{eq:definition_of_R}
\end{equation}
At $R=1$, $h \to -i\pi/6$ meaning that the field term is unitary. Away from this point, both field and interaction terms are non-unitary. The phase diagram for these parameters is shown in Figure \ref{fig:phase_diagram}.

Of particular interest is the time crystal phase, also known as the non-ferromagnet (NFM1) in \cite{PhysRevResearch.4.013018}. In the thermodynamic limit, this phase shows power-law decaying 2-time correlations of the $z$-magnetization, 
\begin{align}\label{CN}
C(N) &= \langle \sigma_{j}^z (N)\sigma_{j}^z (0) \rangle =  \frac{\mathrm{Tr}[\sigma_j^z (\mathcal U^{\dagger})^N \sigma_{j}^z \rho_0 \mathcal U^N]}{\mathrm{Tr}[(\mathcal U^{\dagger})^N \rho_0 \mathcal U^N]},
\end{align}
which is identified as quasi-long-range order in time. Since non-Hermitian time evolution pushes towards the steady state, we are free to choose an arbitrary initial state $\rho_0$ as long as we are interested in late-time non-transient behavior. We choose to average over a random ensemble of initial states, which is equivalent to choosing $\rho_0 \propto I$. In this paper, we specifically select site $1(j=1)$.

\subsection{Quantum circuit with post-selection}

The gate-based structure of this Floquet model suggests to consider its implementation on a quantum computer. However, it is challenging to realize the non-Hermitian terms, as these are not simply done via unitary gates. A number of different possibilities for realizing non-Hermitian time evolution have been proposed in recent years \cite{PhysRevB.105.054304,PRXQuantum.3.010320,PRXQuantum.2.010342,Motta2020,shen2023observationnonhermitianskineffect,Leadbeater_2024}. For this work, we choose the conceptually simplest path, namely  coupling the system to an ancilla qubit, followed by measurement and post-selection. This is similar to how non-Hermitian dynamics are realized in many recent experiments \cite{Li2019,PhysRevLett.127.140504} as well as the non-Hermitian evolution that occurs during stochastic Shr\"odinger equation simulations \cite{PhysRevLett.68.580}. Subsequently, we will investigate how realistic noise influences the dynamics of the post-selected system.

\begin{figure}
  \includegraphics[width=0.9\columnwidth]{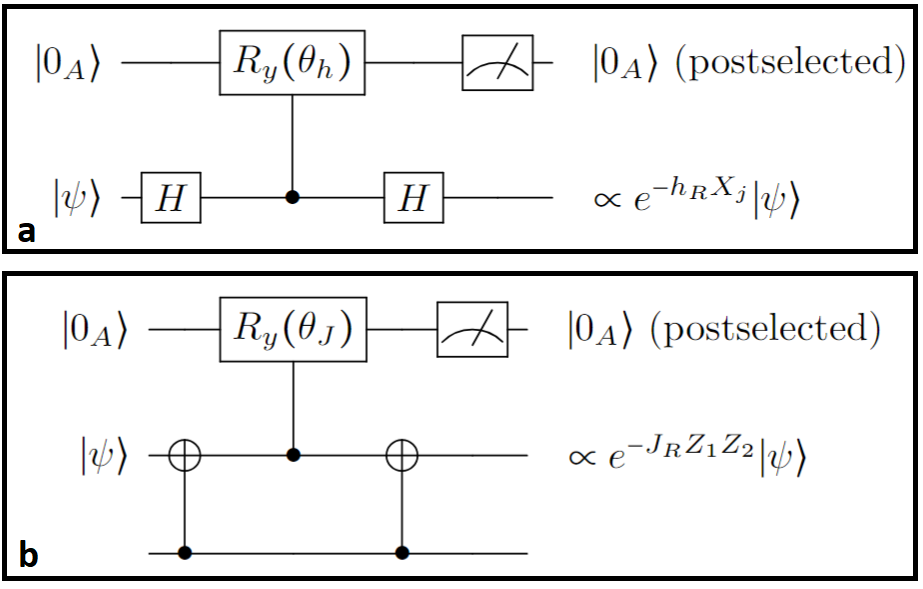}
  \caption{Quantum circuits to realize imaginary time evolution. The field $h_R$ and angle $\theta_h$ are related by $\theta_h = 2\cos^{-1} \left(e^{2h_R}\right)$, and similarly for $J_R$ and $\theta_J$.}
  \label{fig:circuits}
\end{figure}

The Hamiltonian generating each gate can be broken into a sum of Hermitian and anti-Hermitian parts, giving unitary and non-unitary dynamics, respectively. The Hermitian dynamics is easily realized via unitary gates. Therefore, we focus our discussion on realizing the non-unitary gates, such as $e^{-h_R X_j}$. The circuit for doing so is shown in Figure \ref{fig:circuits}a. 
The key component of this circuit is the controlled-$y$-rotation gate implemented with qubit $j$ as the control and an ancilla in state $|0\rangle$ as the target. The controlled rotation entangles the system qubit and ancilla such that the ancilla state depends on the initial $z$-eigenvalue of the system qubit. Therefore, a measurement of the ancilla corresponds to a certain weak measurement of $Z_j$. Averaging over the measurement outcomes is equivalent to the action of pure dephasing, while post-selecting on the outcome $|0_A\rangle$ results in the non-Hermitian gate $e^{-h_R Z_j}$. The rotation angle $\theta_h = 2\cos^{-1} \left(e^{2h_R}\right)$  will reproduce the field $h_R$, as derived in Appendix \ref{sec:app_post_selecation}. Finally, the $Z$ gate is rotated to an $X$ gate via Hadamard rotations.

The above procedure is applied for each of the field terms separately. The bond term, $e^{-J_R Z_1 Z_2}$ , is accomplished by a similar post-selected circuit, shown in Figure \ref{fig:circuits}b. CNOT gates are used to map the value of the parity, $Z_1 Z_2$, to the state of qubit 1. Together with the unitary parts of the gates, these constitute one Floquet cycle, which is repeated $N$ times to get $\mathcal U^N$. The circuit for one Floquet cycle acting on our two-site system is shown in Fig.~\ref{fig:totalcir}. In practice, a further transpilation into experimentally available gates is required, details of which are provided in Appendix \ref{sec:app_decomposition}.

\subsection{Two-time correlation function}
\label{sec:two-time-correlation-function}

As in \cite{PhysRevResearch.4.013018}, the observable of interest will be the infinite temperature two-time autocorrelation function, Eq.~\ref{CN}. A few different methods exist for obtaining two-time correlations on a quantum computer, such as the Hadamard test \cite{Kokcu2024}, which adds another ancilla and thus, potentially, additional complications. The fact that we are working at infinite temperature allows a different option, namely random sampling over basis states. Specifically, writing the trace as a sum over all $z$-basis states $|\psi_{i}\rangle$, the correlation function becomes a sum over expectation values:
\begin{align}\label{MC}
    C(N) &= \frac{\sum_{i} \langle \psi_i|\sigma_j^z (\mathcal U^{\dagger})^N \sigma_{j}^z \mathcal U^N|\psi_i\rangle}{\sum_{i}\langle\psi_{i}|(\mathcal U^{\dagger})^N \mathcal U^N|\psi_{i}\rangle} \\ 
    &= \frac{\sum_{i} s_{ij} \langle \psi_i (N)| \sigma_{j}^z |\psi_i (N) \rangle}{\sum_{i}\langle\psi_{i}(N)|\psi_{i}(N)\rangle},
\end{align}
where $\sigma_j^z|\psi_i\rangle = s_{ij} |\psi_i\rangle$ and $|\psi_i(N)\rangle=\mathcal U^N |\psi_i\rangle$ is the unnormalized state obtained after $N$ Floquet cycles, with normalization $\mathcal{N}_i(N) = \langle \psi_i(N) | \psi_i(N) \rangle$. If $|\phi_i(N)\rangle = \mathcal{N}_i^{-1/2}|\psi_i(N)\rangle $ is the normalized state that one actually has access to, then $C(N)$ can be rewritten
\begin{equation}
     C(N) = \frac{\sum_{i} s_{ij} \mathcal{N}_i(N) \langle \phi_i (N)| \sigma_{j}^z |\phi_i (N) \rangle}{\sum_{i} \mathcal{N}_i(N)}.
\end{equation}

\begin{figure*}
  \includegraphics[width=0.65\textwidth]{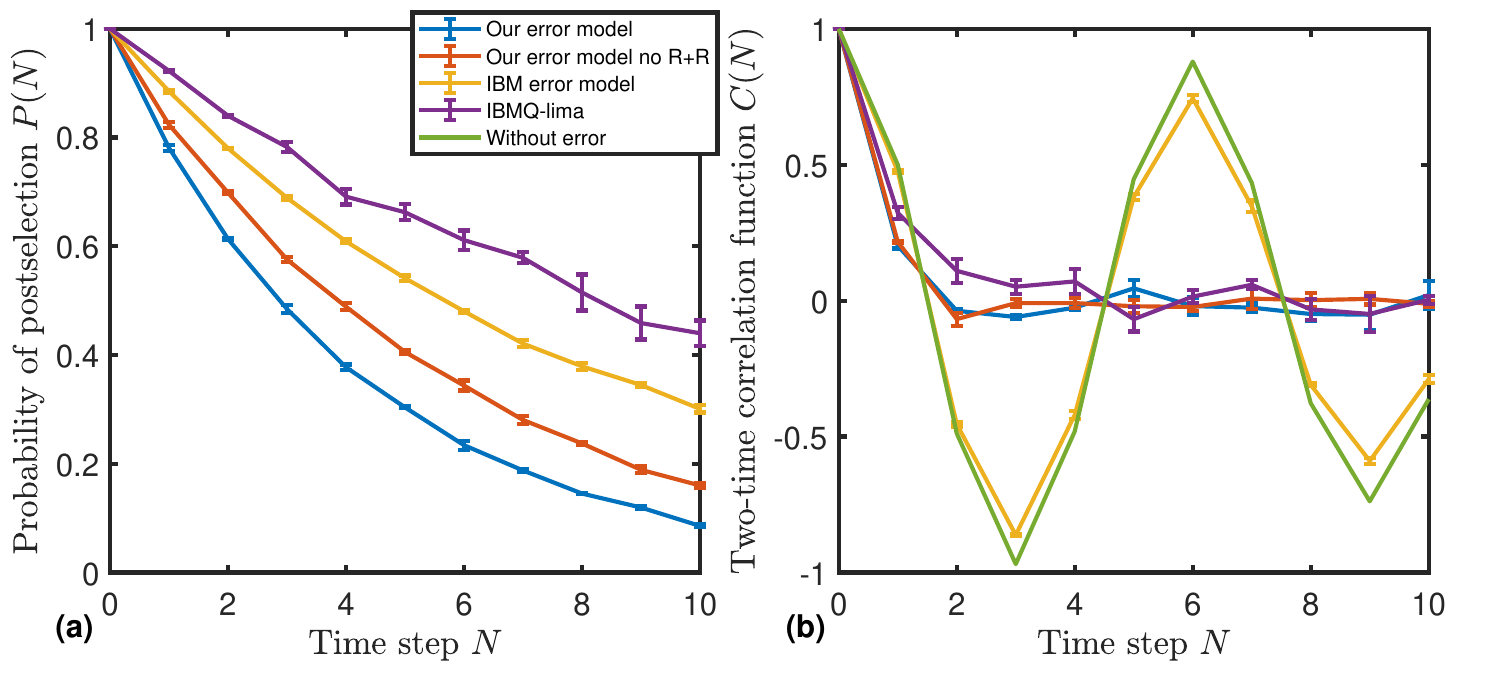}
  \caption{(a) Postselection probability and (b) two-time correlation function comparing experimental data from ibmq-lima to various noise models and exact diagonalization in the absence of noise. Parameter is chosen as $R=1$, i.e., $\tanh \beta = e^{i\pi/3}$. The IBM error model, obtained from Qiskit, is intended to directly model ibmq-lima. Single qubit gates take time $T_{\text{single}}=3.55\times 10^{-8}$ s, 2-qubit gates take $T_{\text{double}}=6.6\times 10^{-7}$ s, reset gates and measurements take $T_{\text{reset}}=1.31\times 10^{-6}$ s and $T_{\text{measure}}=1.4\times 10^{-6}$ s. Noise parameters are $T_1=4.128\times 10^{-4}$ s, $T_2=2.752\times 10^{-4}$ s, depolarizing probability $p_{\text{single}}=0.0001$ and $p_{\text{double}}=0.004$ for 1- and 2-qubit gates, and readout error $p_{\text{read}}=0.02$. We find much better fit to experimental data by modifying parameters to obtain our noise model, with $T_1=1.651\times 10^{-5}$ s, $T_2=1.101\times 10^{-5}$ s, $p_{\text{single}}=0.0025$, and $p_{\text{double}}=0.1$. We also see that removing readout and reset (R+R) error has little effect.}
  \label{fig:ED_lima}
\end{figure*}

Since the postselection procedure gives a normalized quantum state experimentally, we need to employ a separate method to calculate the normalization factor. For initial state $|\phi_{i,0}\rangle = |\phi_{i}\rangle = |\psi_{i}\rangle$, we define
\begin{equation}
\begin{aligned}\label{normalization}
     \mathcal{W}|\phi_{i,0}\rangle &\equiv \mathcal{\eta}_{i,1}|\phi_{i,1}\rangle\\
     \mathcal{V}|\phi_{i,1}\rangle &\equiv \mathcal{\eta}_{i,2}|\phi_{i,2}\rangle\\
     \mathcal{W}|\phi_{i,2}\rangle &\equiv \mathcal{\eta}_{i,3}|\phi_{i,3}\rangle\\
     \mathcal{V}|\phi_{i,3}\rangle &\equiv \mathcal{\eta}_{i,4}|\phi_{i,4}\rangle\\
    &\vdots\\
     \mathcal{W}|\phi_{i,2N-2}\rangle &\equiv \mathcal{\eta}_{i,2N-1}|\phi_{i,2N-1}\rangle\\
     \mathcal{V}|\phi_{i,2N-1}\rangle &\equiv \mathcal{\eta}_{i,2N}|\phi_{i,2N}\rangle\\
\end{aligned}
\end{equation}
where $\eta_{i,j}$ is the normalization factor required to give normalized states $|\phi_{i,j}\rangle$ at each step $j = 1,2,\cdots,2N$. The normalized states $|\phi_{i,j}\rangle$ are what come out from the quantum circuit; in particular $|\phi_{i,2N}\rangle = |\phi_{i}(N)\rangle$. Putting the steps together, we get
\begin{equation}\label{eq:N}
\begin{aligned}
    \mathcal U^N |\psi_i\rangle &= \prod_{j=1}^{2N} \mathcal{\eta}_{i,j} |\phi_{i,2N}\rangle\\
    \mathcal{N}_{i}(N) &= \prod_{j=1}^{2N} |\mathcal{\eta}_{i,j}|^2
\end{aligned}
\end{equation}
Then the individual terms $\eta_{i,j}$ can be obtained in terms of expectation values by expanding out the exponential. For instance, for $j$ odd,
\begin{align*}
    |\mathcal{\eta}_{i,j}|^2 & = \langle \phi_{i,j-1}| e^{2 J_R Z_1 Z_2}|\phi_{i,j-1}\rangle \\
    & = \cosh [2J_R]  + \sinh [2J_R] \langle \phi_{i,j-1}|Z_1 Z_2|\phi_{i,j-1}\rangle.
\end{align*}
Thus, we stop the circuit after each non-unitary gate and measure these observables in order to determine the normalization factor. Note that the normalization is directly tied to the probability of post-selection 
\begin{equation*}
P(N) = |e^{-h}|^{4N}|e^{-J}|^{2N} \mathcal{N}(N),    
\end{equation*}
as shown in Appendix \ref{sec:app_post_selecation}.

\section{Results}
\label{results}

The above circuits were implemented in IBM Qiskit. For such small system sizes, exact results can be obtained using exact diagonalization (ED) of the non-Hermitian Floquet circuit, whereas the Qiskit simulation used simulated circuit sampling as described above. As an initial confirmation of the methods, e.g., the two-time correlation function simulation from Section \ref{sec:two-time-correlation-function}, we compared Qiskit simulations to ED and confirmed matching (see Appendix \ref{sec:app_decomposition}).

The real question of interest is how robust these techniques are to experimental noise. To begin answering this question, we collected data from a real quantum device -- ibmq-lima, a.k.a. the quantum processing unit (QPU) -- which are shown in Fig.~\ref{fig:ED_lima}. The data illustrate that the two-time correlation function decreases rapidly to zero and does not show notable oscillations. This is true even if we change microscopic parameters, such as $R$. On its face, this is not too surprising, as today's quantum computers are still quite noisy, which will clearly degrade the response. However, it is not obvious whether these oscillations will be as sensitive to noise as those in a unitary time crystal because the noiseless non-Hermitian dynamics is contracting, similar to Lindblad time evolution, leading towards a single steady state that should be less sensitive to weak perturbations than unitary dynamics. We expect these non-Hermitian dynamics to compete with the open system dynamics from the bath, leading to a steady state that may or may not be smoothly connected to the one obtained without noise.

To investigate the observed absence of oscillation, we obtained an error model for ibmq-lima directly from Qiskit, which is intended to simulate the real device. The default error model considers five sources of error: depolarization, thermal relaxation, dephasing, measurement error and reset error, with rates and probabilities determined by the ibmq-lima hardware.
Surprisingly, when these parameters are used in the Qiskit simulator, the results do not match well at all with data from ibmq-lima, as seen in Figure \ref{fig:ED_lima}. Indeed, with the noise parameters given, the ibmq-lima simulator shows little difference from the noiseless case. This mystery suggests potential miscalibration of the device and/or inaccurate modeling of device errors, which are crucial for future development on ibmq-lima and other similar quantum devices. However, in order to maintain focus on the role of non-Hermitian dynamics and post-selection on the results from the noisy quantum computer, we defer further investigation of the precise origin of this difference between simulator and experiment to future work.

Our work will instead focus on a theoretical attempt to understand stability of the oscillatory behavior in more general noisy devices. As a first test of noise-dependence, we tune the parameters in the error model to attempt to better fit the data from the QPU. As shown in Fig.~\ref{fig:ED_lima}, the data is much better fit by significantly increasing the noise strength in the model (blue and red curves), though we note that the final values obtained suggest that ibmq-lima is much noisier than most modern quantum computers. By comparing the results with reset and measurement errors turned on and off (blue and red curves), we also find that these errors are not dominant in our results. Therefore, in the remainder of this paper we will restrict ourselves to three of the most important sources of noise, namely depolarization, relaxation, and dephasing. In the section below, we will see that, for each of these noise channels, oscillations do not survive, which we confirm by both calculation of two-time correlations and superoperator spectrum. 


\subsection{Noise channels}

We start by introducing details of the three noise channels we consider and discuss how to think of them in the language of Floquet superoperators. The first error channel we consider is depolarizing error, which results in the quantum channel $\mathcal{E}(\rho) = \sum_i K_i \rho K_i^\dagger$ with Kraus operators \cite{PhysRevA.104.062432}
\begin{equation}
    \begin{aligned}
        K_{D0} = \sqrt{1-p} I,\enskip K_{D1} = \sqrt{\frac{p}{3}} X\\
        K_{D2} = \sqrt{\frac{p}{3}} Z,\enskip K_{D3} = \sqrt{\frac{p}{3}} Y\\
    \end{aligned}
\end{equation}
Vectorizing the density matrix, $\rho \to |\rho\rrangle$, we get the superoperator 
\begin{align}
    \mathcal{E}(\rho) &= \mathcal{K}|\rho\rrangle
\end{align}
where
\begin{align}
    \mathcal{K} = 
    \begin{pmatrix}
        1 - \frac{2}{3}p & 0 & 0 & \frac{2}{3}p\\
        0 & 1-\frac{4}{3}p & 0 & 0\\
        0 & 0 & 1-\frac{4}{3}p & 0\\
        \frac{2}{3}p & 0 & 0 & 1 - \frac{2}{3}p
    \end{pmatrix}
\end{align}

In our error model, depolarizing errors occur every time a gate is applied, all sharing identical parameter $p$. For two-qubit gates, depolarizing errors affect both qubits independently.

\begin{figure*}
  \includegraphics[width=0.65\textwidth]{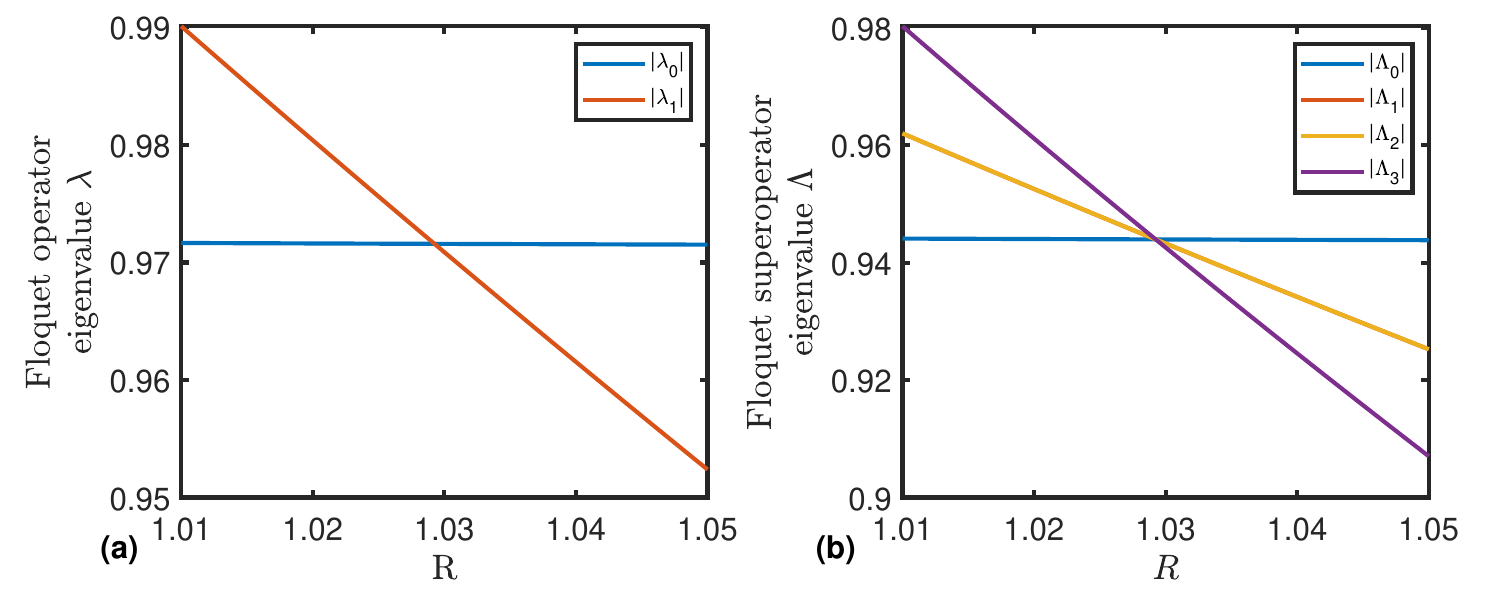}
  \caption{Largest magnitude eigenvalues of (a) the Floquet operator $\mathcal{U}$ and (b) superoperator $\mathcal{K}_f$ as a function of $R=|\tanh \beta|$ near $R_{c2}=1.029$ in the absence of noise. The superoperator eigenvalues can be obtained from pairwise products of the operator eigenvalues and their complex conjugates: $\Lambda_{i,j} = \lambda_i \lambda_j^\ast$ where the index $(i,j)$ is subsequently parsed into a single label $0$ to $3$. \label{fig:operator_vs_superoperator_eigvals}}
\end{figure*}

The second error types we consider are relaxation and dephasing, which are characterized by the characteristic decay time of population ($T_1$) and coherence ($T_2$). Thermal relaxation and dephasing errors  occur concurrently, as energy decay influences the dephasing process. If $T_1\ge T_2$, that is, $\frac{1}{T_2} = \frac{1}{T_1} + \frac{1}{T_\varphi}$ where $T_\varphi$ is the rate of pure dephasing, then the superoperator representing the combined noise channel is 
\begin{equation}
    \mathcal{K} = 
    \begin{pmatrix}
        p_I + p_Z + p_{\mathrm{reset}}p_0 & 0 & 0 & p_{\mathrm{reset}}p_0\\
        0 & p_I - p_Z & 0 & 0\\
        0 & 0 & p_I - p_Z & 0\\
        p_{\mathrm{reset}}p_1 & 0 & 0 & p_I + p_Z + p_{\mathrm{reset}}p_1
    \end{pmatrix}
\end{equation}
where $ p_{\mathrm{reset}} = 1 - \exp\left(-\frac{T}{T_1}\right)$ is the probability of a $T_1$ decay process during time $T$, $p_0=1-p_1$ is the equilibrium thermal occupation of the qubit ground state, $p_Z = \frac{1}{2}(1 - p_{\mathrm{reset}})\left(1-e^{-T(\frac{1}{T_2}-\frac{1}{T_1})}\right)$ is the probability of dephasing, and $p_I=1 - p_Z - p_{\mathrm{reset}}$ is added to maintain normalization. 
A similar expression can be derived for the case $T_1<T_2\le 2 T_1$, as shown in Appendix \ref{sec:choi_noise_limit}.

\subsection{Post-selection}

In addition to these non-unitary superoperators due to noise, the quantum circuit consist of unitary quantum gates and post-selected measurements. Unitaries are trivially written as superoperators $U \rho U^\dagger \to U^\ast \otimes U |\rho\rrangle \equiv \mathcal{K}_\mathrm{gate}|\rho\rrangle$.
Measurement can be written as a sum over projectors onto the two possible outcomes. To get post-selection, we simply ignore the undesired outcome. This gives a superoperator
\begin{align}
    \mathcal{K}_{\mathrm{post}} &= \Pi_\mathrm{post}^* \otimes \Pi_\mathrm{post}
\end{align}
where $\Pi_\mathrm{post}^2 = \Pi_\mathrm{post}$ is a projector onto the post-selected subspace. Combining the effect of these unitary gates and non-unitary noise + post-selection superoperators, we get a combined superoperator equal to
\begin{align}
    \mathcal{K}_f &= \mathcal{T} \prod_i \mathcal{K}_{i}
\end{align}
where $i$ represents different operations and $\mathcal{T}$ represents time-ordering. If $i$ corresponds to a quantum gate, then $\mathcal{K}_{i} = \mathcal{K}_{\mathrm{thermal},i}  \mathcal{K}_{\mathrm{depolarizing},i} \mathcal{K}_{\mathrm{gate},i}$. If $i$ corresponds to measurement and postselection, then $\mathcal{K}_{i} = \mathcal{K}_{\mathrm{post}}$.

\section{Results of error on post-selected dynamics}
\label{s:result_anlytical}

In this section, we explore the impact of independently introducing $T_1$, $T_2$, and decoherence on the analytical results. Specifically, we examine whether these factors preserve or disrupt the system's degeneracy. In the absence of degeneracy, the system's behavior is primarily determined by the eigenvalue with the smallest decay rate. Conversely, when the system exhibits degeneracy — characterized by the largest eigenvalues of the Floquet superoperator having equal magnitude — it will oscillate among these degenerate states.

\begin{figure*}
  \includegraphics[width=0.9\textwidth]{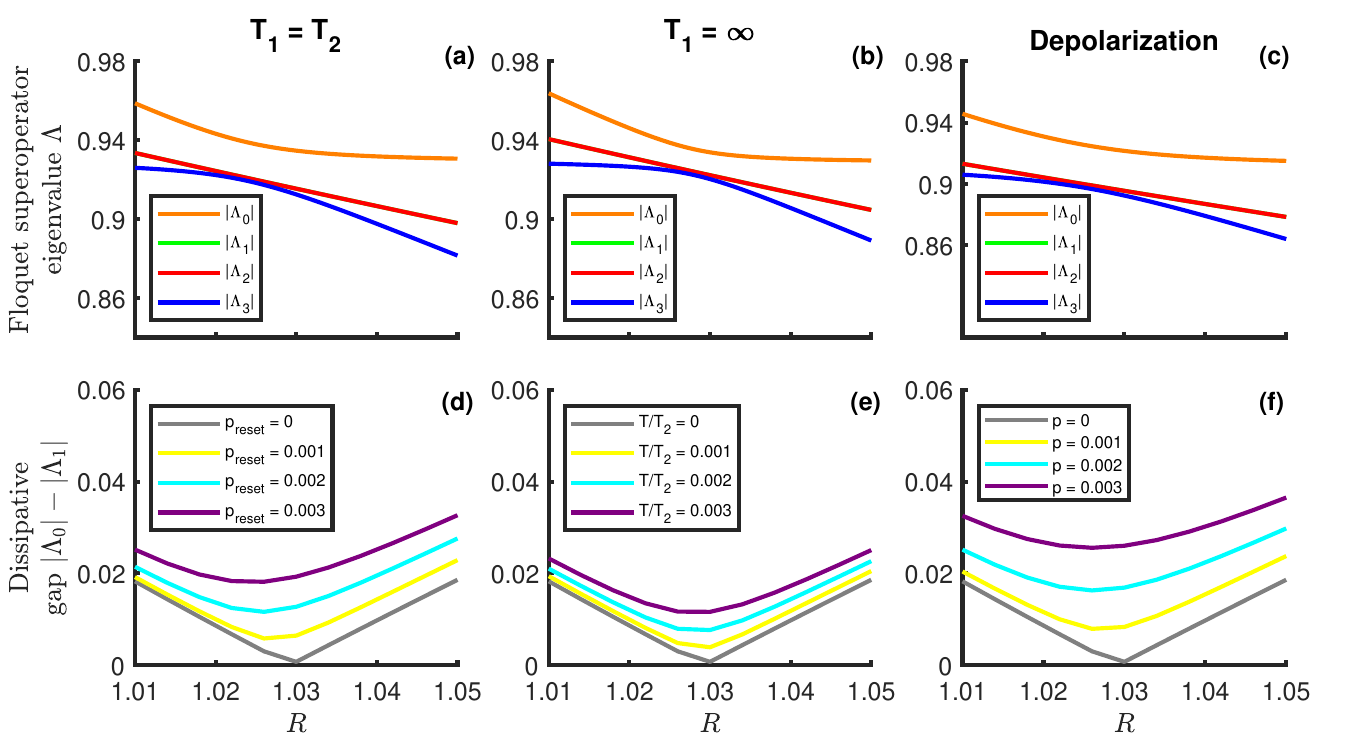}
  \caption{Noise-dependence of (a-c) leading Floquet superoperator eigenvalues and (d-f) dissipative gap. Data is taken for the same radial cut as Figure \ref{fig:operator_vs_superoperator_eigvals}. In all cases, $|\Lambda_1| = |\Lambda_2|$. Noise parameters for the top panels are (a) $p_{\text{reset}} = 1-\exp{\left(-\frac{T}{T_1}\right)} = 0.003$, (b) $\frac{T}{T_2}=0.003\approx 2p_Z$, and (c) $p=0.003$.}
  \label{fig:superoperator_error}
\end{figure*}

\subsection{Evolution operator and superoperator without error}

When the quantum circuit doesn't have noise, the non-unitary Floquet operator $\mathcal U$ and the superoperator $\mathcal{K}_f$ are related by $\mathcal{K}_f=\mathcal U^{*}\otimes \mathcal U$, where $\rho_f = \mathcal U\rho \mathcal U^{\dagger}$. Writing $\lambda_0$ and $\lambda_1$ for the two eigenvalues of $\mathcal U$ with the largest magnitudes,  the four most relevant eigenvalues of the superoperator $\mathcal{K}$ are $|\lambda_0|^2, \lambda_0^*\lambda_1, \lambda_1^*\lambda_0$ and $|\lambda_1|^2$. At points where $|\lambda_0|$ and $|\lambda_1|$  all four of the superoperator eigenvalues become degenerate in magnitude. For our choice of parameters, the two-fold degeneracy of $\mathcal U$ occurs at $R_{c1}=0.974$ and $R_{c2}=1.029$; we focus on $R_{c2}=1.029$. When $R = R_{c2}$, coherences between these degenerate states will cause oscillations at a rate equal to $\mathrm{arg}\left( \lambda_0^*\lambda_1 \right)$, leading to long-lived oscillations. Away from this point, these oscillations will be damped by a factor of $\mathrm{min}\left[|\lambda_0/\lambda_1|^2,|\lambda_1/\lambda_0|^2\right]$ per Floquet cycle. An example of these spectra are shown in Figure \ref{fig:operator_vs_superoperator_eigvals}.

\subsection{Superoperator with error}

The next step is to establish whether the degeneracy persists when noise is present. When noise is introduced, we cannot describe the system using the Floquet operator; however, the superoperator can still effectively describe the system. Figure \ref{fig:superoperator_error} shows how the four slowest-decaying eigenvalues of the superoperator are affected by noise. 

Specifically, we consider noise in three separate limits. First we consider the relaxation-dominated limit  $T_1 = T_2$, which eliminates $p_Z$. The results are shown in Fig.~\ref{fig:superoperator_error}. The degeneracy is clearly broken, favoring a unique steady state. Gaps seem to scale linearly as $|\Lambda_0|-|\Lambda_1| \sim p_\mathrm{reset}$ for small $p_\mathrm{reset}$, suggesting that first-order degenerate perturbation theory should be sufficient to capture the physics. 

Second, we consider pure dephasing by setting $T_1 = \infty$. The results, shown in Fig.~\ref{fig:superoperator_error}, are similar to the case with $T_1=T_2$. Again the gap opens up as $|\Lambda_0|-|\Lambda_1| \sim p_Z$. Finally, we consider the case with only depolarizing error, as illustrated in Figure \ref{fig:superoperator_error}. The results are similar and, again, consistent with having first-order corrections dominate.


\subsection{Perturbative treatment}

We have found that, for each of these noise sources treated separately, degeneracy is lost at arbitrarily weak noise and the long-lived oscillations cease to exist. In this section, we argue that this will be the case more generally for any combination of these terms.

The key to our argument is that all three noise terms produce a gap which is dominated by first-order degenerate perturbation theory. Therefore, for a weak perturbation $\epsilon$ (a proxy for $p_\mathrm{reset}$, $p_Z$, or $p$), we can consider the degenerate point $R_c$ and linearize the effect of the noise channel. For a single noise channel, we can write it as
\begin{equation}
\mathcal{K}(R_c,\epsilon) \approx \mathcal{K}(R_c,0) + \epsilon \Delta K(R_c,0) \equiv \left[1 + \epsilon \mathcal{L}_\mathrm{eff} \right] \mathcal{K}(R_c,0),
\end{equation}
where we used the fact that commutators are irrelevant at linear order to factor out the effect of noise. The effective Liouvillian for noise can be readily obtained in the eigenbasis of $\mathcal{K}(R_c,0)\equiv \mathcal K_0=\mathcal U^{*}(R_c)\otimes \mathcal U(R_c)$. We write the eigenmodes of $\mathcal{K}_0$ as $|\Lambda_j\rrangle$, and use the notation $\llangle \Lambda_j|$ to denote their Hermitian conjugates which, in general, are not the left eigenmodes. Then we see that $\llangle \Lambda_i|\mathcal{K}(R_c,\epsilon)|\Lambda_j\rrangle \approx \Lambda_{j} \delta_{ij} + \epsilon \Lambda_j \llangle \Lambda_i|\mathcal{L}_\mathrm{eff}|\Lambda_j\rrangle$. Thus we can write 
\begin{align}
    \mathcal{L}_\mathrm{eff}^{ij} = \lim_{\epsilon \to 0} \frac{\llangle \Lambda_i|\mathcal{K}(R_c,\epsilon)|\Lambda_j\rrangle-\Lambda_j\delta_{ij}}{\epsilon\Lambda_j}
\end{align}

\begin{figure}
  \includegraphics[width=\columnwidth]{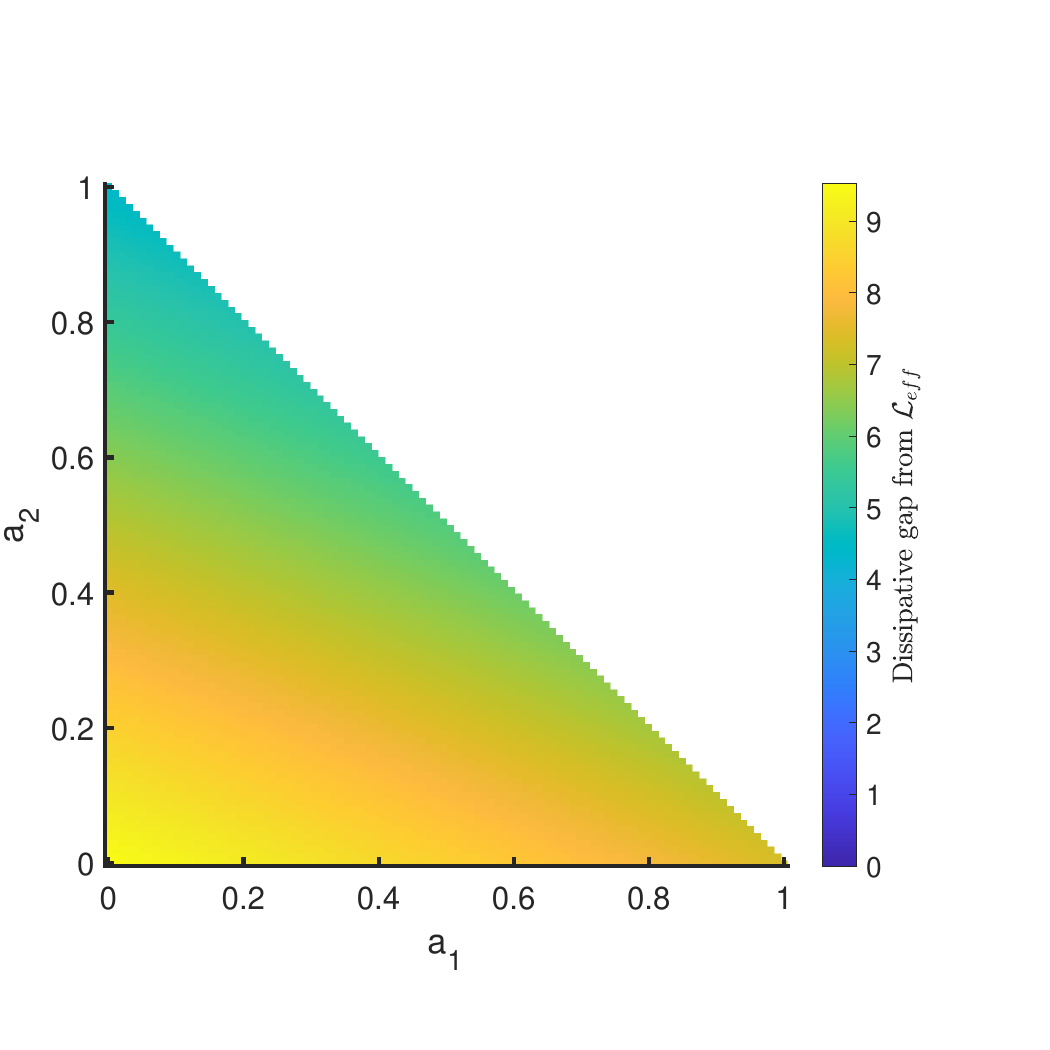}
  \caption{Difference of the magnitude of leading eigenvalues of $\mathcal{L}_\mathrm{eff}$ -- proportional to the dissipative gap at $R_{c2}$ -- as a function of relaxation strength ($a_1$) and dephasing strength ($a_2$). Depolarization strength is set by $a_3 = 1 - a_1 - a_2$, limiting the parameters to $a_1+a_2 \leq 1$ as shown.}
  \label{fig:perturbation}
\end{figure}

This is the matrix that determines the gap structure of the system at small $\epsilon$. In particular, projecting onto the two-dimensional degenerate subspace $|\lambda_0\rangle\langle\lambda_0|$ and $|\lambda_1\rangle\langle\lambda_1|$ we would need $\mathcal{L}_\mathrm{eff} \propto I$ in order to stabilize oscillations. The clearly does not happen for pure relaxation, dephasing, or depolarization, which we label $\mathcal{L}_1$, $\mathcal{L}_2$, and $\mathcal{L}_3$ respectively. However, because their commutator is irrelevant at linear order, we can consider the leading perturbative effect of an arbitrary error channel: $\mathcal{L}_\mathrm{eff} = a_1 \mathcal{L}_1 + a_2 \mathcal{L}_2 + a_3 \mathcal{L}_3$, with $a_j \in [0,1]$ and $\sum_j a_j = 1$ arbitrarily setting the normalization. As seen in Fig.~\ref{fig:perturbation}, none of these combinations successfully eliminated the Lindbladian gap, confirming that our system loses the long-lived oscillations for arbitrary noise within this family.

\section{Discussion}
\label{s:discussion}

We implement the two-qubit version of a non-Hermitian Floquet transverse-field Ising model on a quantum computer and calculate the correlation function in its time crystalline phase. We observe that the expected oscillations are absent on an actual device. We model our system as an open Floquet system and are able to match the observed data to a reasonable degree, although we find that the parameters supplied by Qiskit do not accurately model the real quantum device. We proceed to theoretically analyze the degeneracies of the system's superoperator, taking into account depolarization, relaxation, and dephasing. Our findings reveal that any of these errors individually or together eliminate these degeneracies and therefore damp the oscillations. As the complex gaps are opened generically at first order in perturbation theory, this finding  should generalize to similar cases involving a larger, finite number of qubits, wherein degeneracies occur only at fine-tuned points.

A key point that this work helps clarify is the relationship between degenerate subspaces in non-Hermitian dynamics and decoherence-free subspaces in open quantum systems. Both of these cases involve pure state dynamics in the steady state subspace, but behave differently outside of this subspace -- whether or not they maintain purity of pure states, whether or not they are trace preserving, etc. What our work has shown is that these situations also behave identically with regards to their superoperator spectra within the degenerate steady state manifold, again despite behaving different on the excited states. These connections allows us to see why the non-Hermitian dynamics is not stable to noise, since generic noise breaks the degeneracy in the same way as happens for decoherence-free subspaces. For much larger system sizes, this non-Hermitian model has a dissipative ``Fermi surface'' leading to power law relaxation that has no known analogue in open quantum systems. We expect that, despite this gapless continuum in the absence of noise, a gap will still be opened by generic nosie. However, this remains untested and, more broadly, understanding the scaling by which this more complex slow mode structure is affected by noise remains an interesting open question.


\begin{acknowledgments}
We thank Sarang Gopalakrishnan for helpful discussions. Work was performed with support from the National Science Foundation through award numbers DMR-1945529 and MPS-2228725 (MK) and the Welch Foundation through award number AT-2036-20200401 (MK and WX). Part of this work was performed at the Aspen Center for Physics, which is supported by NSF grant number PHY-1607611. We used the computational resources ibmq-lima operated by the IBM. The Flatiron Institute is a division of the Simons Foundation.

\end{acknowledgments}

\appendix
\begin{widetext}

\section{Decomposing and testing quantum circuits}
\label{sec:app_decomposition}

Prior to implementation on a quantum device, the quantum circuits shown in the main text must be combined with the unitary pieces of the time evolution and transpiled to near-term hardware. In this section, we show how that process is done and how we checked these circuit models to verify correctness.

The circuit for one Floquet cycle acting on our two-site system is shown in Fig.~\ref{fig:totalcir}. This is obtained by straightforwardly combining the non-Hermitian gates shown in Figure 2 of the main text with conventional unitary gates to implement the unitary part of the time evolution. To execute this on ibmq-lima, a further transpilation must be done. ibmq-lima supports the basis gate set Identity, Controlled-X, Rotation-Z, and SX. This was done via Qiskit's transpilation tool at optimization level 3. The transpiled circuit is shown in Fig.~\ref{fig:decomposed_circuit}. This transpiled circuit is what we used for all data shown in the paper, including both the QPU data and the classical simulations with or without noise.

\begin{figure*}
\centering
\includegraphics[width=\textwidth]{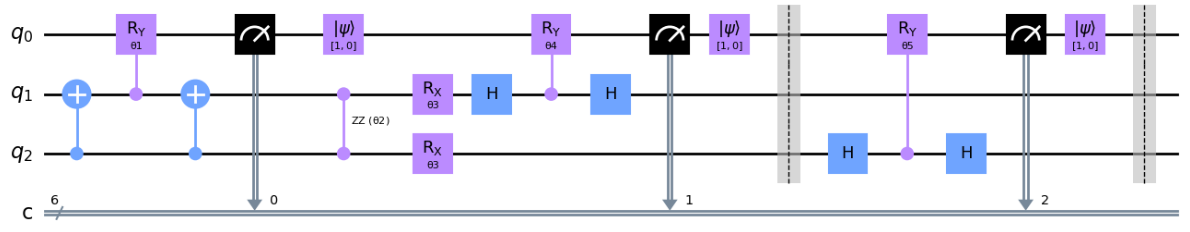}
\caption{Quantum circuit implementing one Floquet cycle of our protocol after postselection on all measurement outcomes to $0$. $q_0$ is the ancilla qubit, while $q_1$ and $q_2$ are the two system qubits.}
\label{fig:totalcir}
\end{figure*}

\begin{figure*}
\centering
\includegraphics[width=\textwidth]{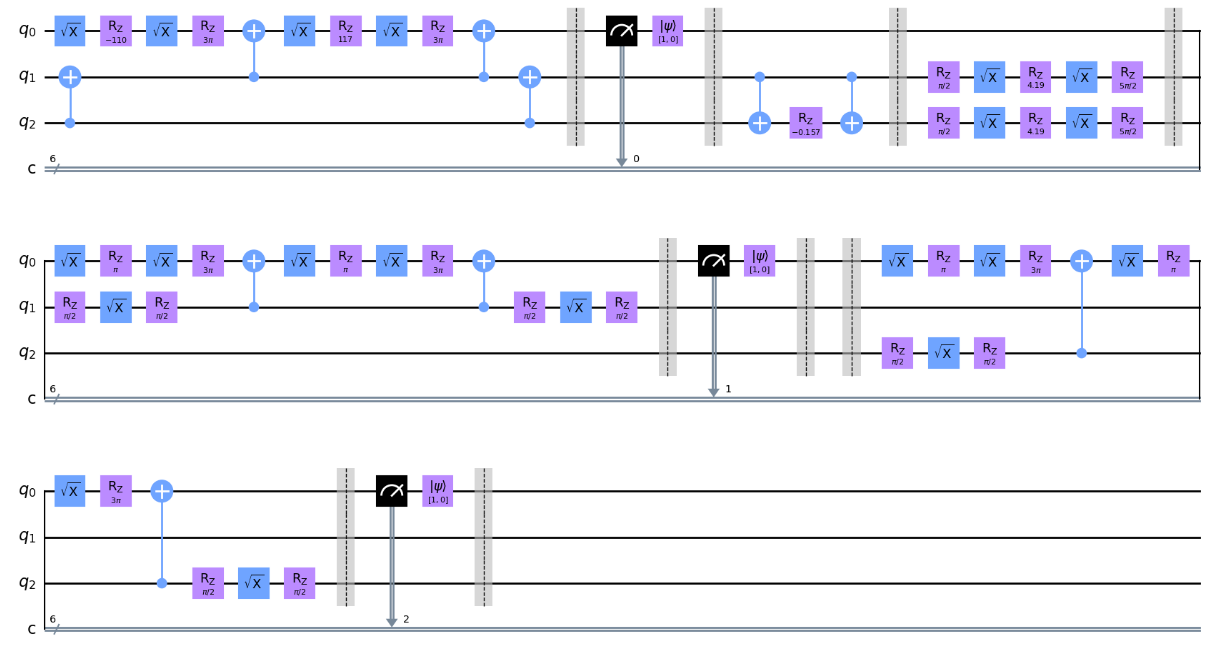}
\caption{Quantum circuit implementing one Floquet cycle of our protocol using gates transpiled for ibmq-lima. While logically equivalently to the circuit from Figure \ref{fig:totalcir}, the gate decomposition shown here is physically implementable on existing hardware. All analysis, including QPU simulations and error modeling on classical hardware, is done for this larger, transpiled circuit.}
\label{fig:decomposed_circuit}
\end{figure*}


We utilized Qiskit to execute our quantum circuit. Initially, to verify convergence of the autocorrelation function calculation, we conducted a simulation of the circuit without noise and then compared the outcomes with those obtained through exact diagonalization (ED). The comparison indicates a match between the results from the simulator and those derived from the ED method, affirming the reliability of the sampling procedure. Additionally, we obtained data regarding the probability of postselection, as depicted in Fig.~\ref{fig:simulator_ED_lima} and Fig.~\ref{fig:simulator_ED_P}. The data reveals that the probability of postselection does not decrease rapidly. This observation suggests the complex gap is relatively small.

\begin{figure*}
  \subfigure[]{\includegraphics[width=0.32\textwidth]{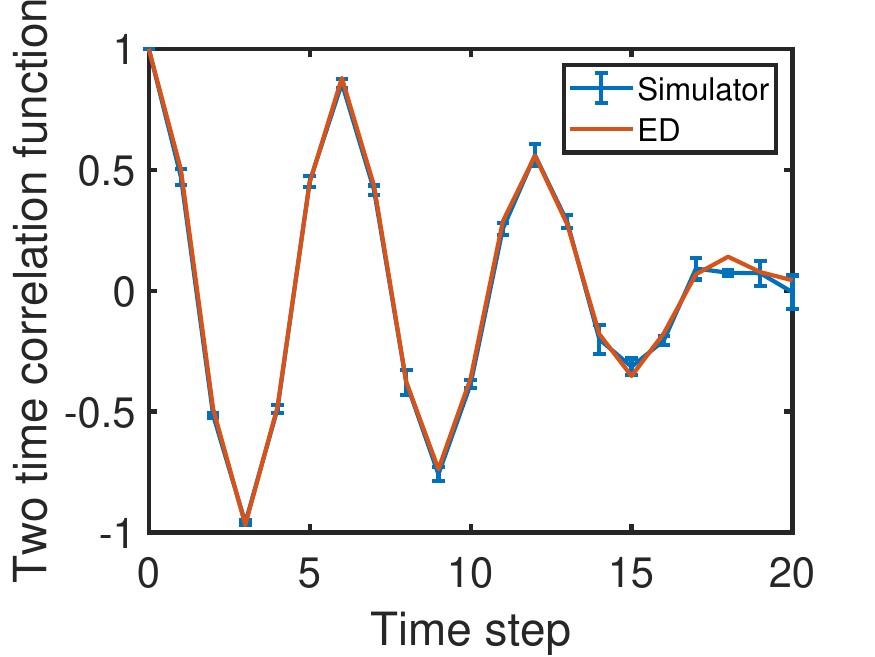}\label{fig:simulator_ED_lima}}\quad
  \subfigure[]{\includegraphics[width=0.32\textwidth]{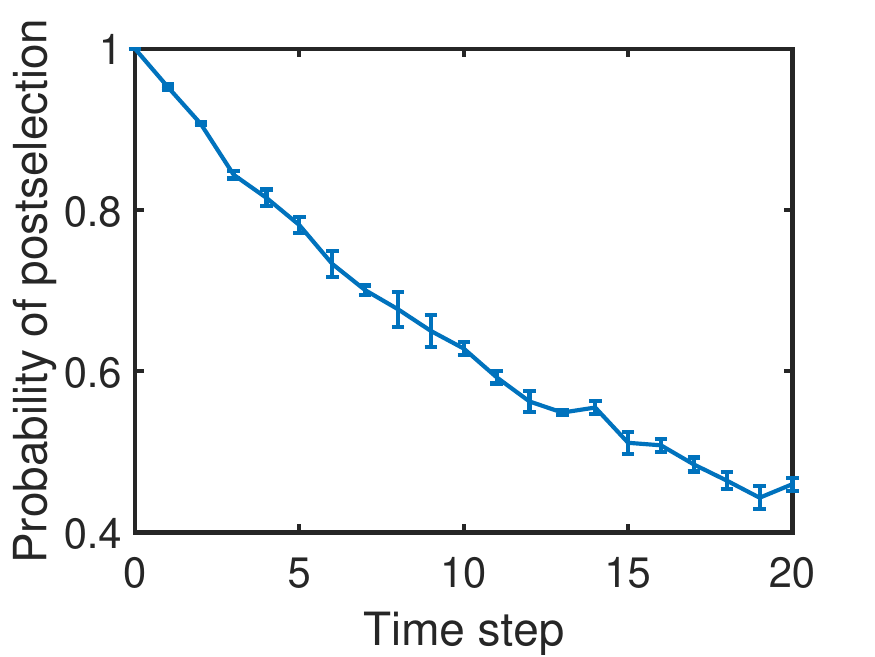}\label{fig:simulator_ED_P}}
  \caption{(Left) Two-time correlation function and (right) post-selection probability in the NFM1 phase comparing a noiseless simulator of the quantum circuit and exact diagonalization. This data is used to verify our sampling technique for initial state averaging of the correlation function. Data is taken for $R=1$, i.e., $\tanh{\beta}=\exp{(i\pi/3)}$.}
\end{figure*}

\section{Details of post-selected non-unitary gates}
\label{sec:app_post_selecation}

In this appendix, we derive the method for doing non-Hermitian dynamics referenced in Figure 2. We start by introducing the circuit for $e^{-h_R Z_j}$ for $h_R\in \mathds{R}$, from which the desired terms are easily obtained. Consider the following quantum circuit:
\[
\Qcircuit @C=0.5em @R=2em {
\lstick{|0_A\rangle} &  \gate{R_y(\theta)}  & \meter & \qw & \rstick{|0_A\rangle \text{ (postselected)}} \\
\lstick{|\psi\rangle} &  \ctrl{-1}  & \qw & \qw & \rstick{\propto e^{-h_RZ_j}|\psi\rangle} 
}
\]
The ancilla is initialized in state $|0_A\rangle$ and, for illustration, we will assume that the qubit on which the gate is applied is in the pure state $|\psi\rangle$. After implementing a controlled y-rotation, we have 
\begin{equation}
    |\phi_f\rangle = \left(\cos{\frac{\theta}{2}}|0_A\rangle + \sin{\frac{\theta}{2}}|1_A\rangle\right) \otimes \frac{1-Z}{2}|\psi\rangle
    + |0_A\rangle \otimes \frac{1+Z}{2} |\psi\rangle
\end{equation}
Then, after measuring the ancilla and postselecting on the outcome $|0\rangle$, we have 
\begin{align}\label{eq:post}
    |\phi_{post}\rangle & \propto \left(\frac{\cos{(\theta/2)}+1}{2} + \frac{1-\cos{(\theta/2)}}{2} Z \right)|\psi\rangle
\end{align}
We know that $e^{-h_R Z} = \cosh{h_R} - Z \sinh{h_R}$, so we can equate the ratio of the identity and $Z$ terms to get $\tanh{h_R} = \frac{\cos{\frac{\theta}{2}}-1}{\cos{\frac{\theta}{2}}+1}$, or equivalently $\theta = 2\cos^{-1} \left(e^{2h_R}\right)$. Notice that only the ratio of the $Z$ and identity terms must match, since a different choice of normalization is chosen for $|\phi_{post}\rangle$ as compared to $e^{-h_{R}Z}|0_A\rangle \otimes |\psi\rangle$.

Going from $e^{-h_R Z_j}$ to $e^{-h_R X_j}$ or $e^{-h_R Z_j Z_{j+1}}$ is straightforwardly accomplished by a local rotation. For the transverse field term, $e^{-h_RX_j}$, we add two Hadamard gates:
\[
\Qcircuit @C=0.5em @R=2em {
\lstick{|0_A\rangle} & \qw &  \gate{R_y(\theta_h)} & \qw  & \meter & \qw & \rstick{|0_A\rangle \text{ (postselected)}} \\
\lstick{|\psi\rangle} & \gate{H} &  \ctrl{-1} & \gate{H}  & \qw & \qw & \rstick{\propto e^{-h_{R}X_j}|\psi\rangle} 
}
\]
This rotates the initial $x$ basis of $|\psi\rangle$ to the $z$-basis, such that the $y$-rotation is controlled by the initial $x$-eigenstate of $|\psi\rangle$. Mathematically, this can simply be written $He^{-h_RZ}H = e^{-h_RX}$. Similarly, a CNOT gate with qubit $j$ as the target:
\[
\Qcircuit @C=0.5em @R=2em {
\lstick{|0_A\rangle} & \qw & \gate{R_y(\theta_J)} & \qw & \meter & \qw & \rstick{|0_A\rangle \text{ (postselected)}} \\
\lstick{|\psi\rangle} & \targ & \ctrl{-1} & \targ & \qw & \qw & \rstick{\propto e^{-J_{R}Z_1 Z_2}|\psi\rangle} \\
\lstick{\quad} & \ctrl{-1} & \qw & \ctrl{-1} & \qw & \qw
}
\]
takes the initial $z$-parity value $ZZ \equiv Z_j Z_{j+1}$ and puts it into the $z$-eigenbasis of qubit $j$. Mathematically, $(CNOT)e^{-h_RZ}(CNOT) = e^{-h_RZZ}$.

Finally, we can show that the post-selection probability $P$ is directly related to the normalization $\mathcal{N}$. Consider one step involving a transverse field term on, say, site $1$: $e^{-h_RZ_1}|\phi_{i,j}\rangle = \mathcal{\eta}_{i,j+1}|\phi_{i,j+1}\rangle$. From Eq.\ref{eq:post}, the unnormalized state after post-selection is
\begin{align}
    \left(\frac{\cos{(\theta_h/2)}+1}{2} + \frac{1-\cos{(\theta_h/2)}}{2} Z_1 \right)|\phi_{i,j}\rangle = \frac{\cos{(\theta_h/2)}+1}{2\cosh{h_R}} e^{-h_RZ_1}|\phi_{i,j}\rangle = \frac{\cos{(\theta_h/2)}+1}{2\cosh{h_R}}  \mathcal{\eta}_{i,j+1}|\phi_{i,j+1}\rangle
\end{align}
The normalization of this state gives us the post-selection probability,
\begin{align}
    P_j = \left\vert\frac{\cos{(\theta_h/2)}+1}{2\cosh{h_R}}\right\vert^2 |\eta_{i,j}|^2.
\end{align}
A similar expression holds for $e^{-h_RZ_2}$ and  $e^{-J_RZ_1Z_2}$. Combining these results, we have the probability of post-selection is
\begin{equation}
\begin{aligned}
    P(N) &= \left\vert\frac{\cos{(\theta_h/2)}+1}{2\cosh{h_R}}\right\vert^{4N} \left\vert\frac{\cos{(\theta_J/2)}+1}{2\cosh{J_R}}\right\vert^{2N} \mathcal{N}(N)\\
    &= |e^h|^{4N} |e^J|^{2N} \mathcal{N}(N)
\end{aligned}
\end{equation}
\section{Choi operator for the case $T_1<T_2\le 2 T_1$ }
\label{sec:choi_noise_limit}

For $T_1<T_2\le 2 T_1$, we can express the noise channel via Choi operator \cite{PhysRevA.104.062432}:
\begin{align}\label{eq:Choi}
    \Lambda &= 
    \begin{pmatrix}
        1 - p_1(1 - \exp\left(-\frac{T}{T_1}\right)) & 0 & 0 & \exp\left(-\frac{T}{T_2}\right)\\
        0 & p_1(1 - \exp\left(-\frac{T}{T_1}\right))& 0& 0\\
        0 & 0 & p_0(1 - \exp\left(-\frac{T}{T_1}\right)) & 0\\
        \exp\left(-\frac{T}{T_2}\right) & 0 & 0 & 1 - p_0(1 - \exp\left(-\frac{T}{T_1}\right))
    \end{pmatrix}
\end{align}
The relationship between the Choi operator and the quantum channel is established by mapping the state transformations within a quantum system:
\begin{align}
    \Lambda &= \sum_{i,j} |i\rangle \langle j| \otimes \mathcal{E}(|i\rangle \langle j |)
\end{align}
\begin{align}
    \mathcal{E}(\rho) = Tr_1 [\Lambda (\rho^T \otimes I)]
\end{align}
Given this information, we can express the superoperator for this noise channel as follows:
\begin{align}
    \mathcal{K} = 
    \begin{pmatrix}
        1 - p_1 p_{reset} & 0 & 0 & (1 - p_1) p_{reset}\\
        0 & E_2 & 0 & 0\\
        0 & 0 & E_2 & 0\\
        p_1 p_{reset} & 0 & 0 & [1 - (1 - p_1) p_{reset}]
    \end{pmatrix}
\end{align}
For computational convenience, we typically express $\mathcal{E}(\rho)$ using Kraus operators

\begin{align}
    \mathcal{E}(\rho) &= \sum_i K_i \rho K_i^{\dagger}
\end{align}
where $K_i = \sqrt{\lambda_i} \Psi_i$, $\lambda_i$ are the eigenvalues of $\Lambda$, and $\Psi_i$ consist of the eigenvectors of $\Lambda$ reshaped as matrices.
\end{widetext}


\bibliography{mycitation}

\end{document}